\documentclass[a]{iucr}              % DO NOT DELETE THIS LINE

\usepackage{siunitx}
\usepackage{graphicx}

     \journalcode{J}              % Indicate the journal to which submitted
\begin{document}                  % DO NOT DELETE THIS LINE

     %-------------------------------------------------------------------------
     % The introductory (header) part of the paper
     %-------------------------------------------------------------------------

     % The title of the paper. Use \shorttitle to indicate an abbreviated title
     % for use in running heads (you will need to uncomment it).

\title{Density-based clustering of crystal orientations and misorientations and the orix python library}
\shorttitle{Clustering crystal (mis)orientations}

     % Authors' names and addresses. Use \cauthor for the main (contact) author.
     % Use \author for all other authors. Use \aff for authors' affiliations.
     % Use lower-case letters in square brackets to link authors to their
     % affiliations; if there is only one affiliation address, remove the [a].
\cauthor[a]{Duncan N.}{Johnstone}{dnj23@cam.ac.uk}{}
\author[a]{Ben H.}{Martineau}
\author[a]{Phillip}{Crout}
\author[a]{Paul A.}{Midgley}
\author[b]{Alexander S.}{Eggeman}

\aff[a]{Department of Materials Science and Metallurgy,
University of Cambridge,
27 Charles Babbage Road,
Cambridge CB3 0FS,
\country{United Kingdom}}
\aff[b]{Department of Materials,
The University of Manchester,
Oxford Road,
Manchester M13 9PL,
\country{United Kingdom}}

     % Use \shortauthor to indicate an abbreviated author list for use in
     % running heads (you will need to uncomment it).

\shortauthor{Johnstone et al.}

     % Use \vita if required to give biographical details (for authors of
     % invited review papers only). Uncomment it.

%\vita{Author's biography}

     % Keywords (required for Journal of Synchrotron Radiation only)
     % Use the \keyword macro for each word or phrase, e.g. 
     % \keyword{X-ray diffraction}\keyword{muscle}


     % PDB and NDB reference codes for structures referenced in the article and
     % deposited with the Protein Data Bank and Nucleic Acids Database (Acta
     % Crystallographica Section D). Repeat for each separate structure e.g
     % \PDBref[dethiobiotin synthetase]{1byi} \NDBref[d(G$_4$CGC$_4$)]{ad0002}

%\PDBref[optional name]{refcode}
%\NDBref[optional name]{refcode}

\maketitle                        % DO NOT DELETE THIS LINE

\begin{synopsis}
Data clustering incorporating symmetry is applied to crystal orientations and misorientations and the orix python library for crystal orientation analysis is introduced.
\end{synopsis}

\begin{abstract}
Crystal orientation mapping experiments typically measure orientations that are similar within grains and misorientations that are similar along grain boundaries. Such (mis)orientation data will cluster in (mis)orientation space and clusters are more pronounced if preferred orientations or special orientation relationships are present. Here, cluster analysis of (mis)orientation data is described and demonstrated using distance metrics incorporating crystal symmetry and the density based clustering algorithm DBSCAN. Frequently measured (mis)orientations are identified as corresponding to grains, grain boundaries or orientation relationships, which are visualised both spatially and in three-dimensional (mis)orientation spaces. A new open-source python library, \textit{orix}, is also reported. 
\end{abstract}

     %-------------------------------------------------------------------------
     % The main body of the paper
     %-------------------------------------------------------------------------
     % Now enter the text of the document in multiple \section's, \subsection's
     % and \subsubsection's as required.

\section{Introduction}\label{sec:introduction}

The distribution of crystal orientations in a polycrystalline material (i.e. crystallographic texture) and characteristic misorientations between neighbouring crystals (i.e. orientation relationships) are affected by material processing and influence material properties \cite{Kocks1998, Sutton2007}. Measuring the local crystal orientation throughout a material is therefore common in modern materials characterisation. Such mapping is usually achieved using scanning diffraction techniques such as: electron backscatter diffraction (EBSD) \cite{Schwartz2009}, scanning electron diffraction (SED) \cite{Zaefferer2000, Rauch2008} and X-ray microLaue diffraction \cite{Ice2009}. These techniques involve using a small (nm - $\mu$m) probe to address numerous locations across the specimen while recording diffraction data at each position. Such data can be used to determine the local crystal \textit{orientation} conventionally defined \cite{Rowenhorst2015} as the passive rotation, $g_{i}$, between the crystal coordinate system, $h_{i}$, and a reference specimen coordinate system, $r$, \cite{Morawiec2004}, i.e.

\begin{equation}
r = g_{i} h_{i}
\end{equation}

\begin{figure}
    \caption{Schematic representation of orientations, $g_{i}$, and misorientations, $m$, as transformations between reference frames.}
    \includegraphics[width=0.9\linewidth]{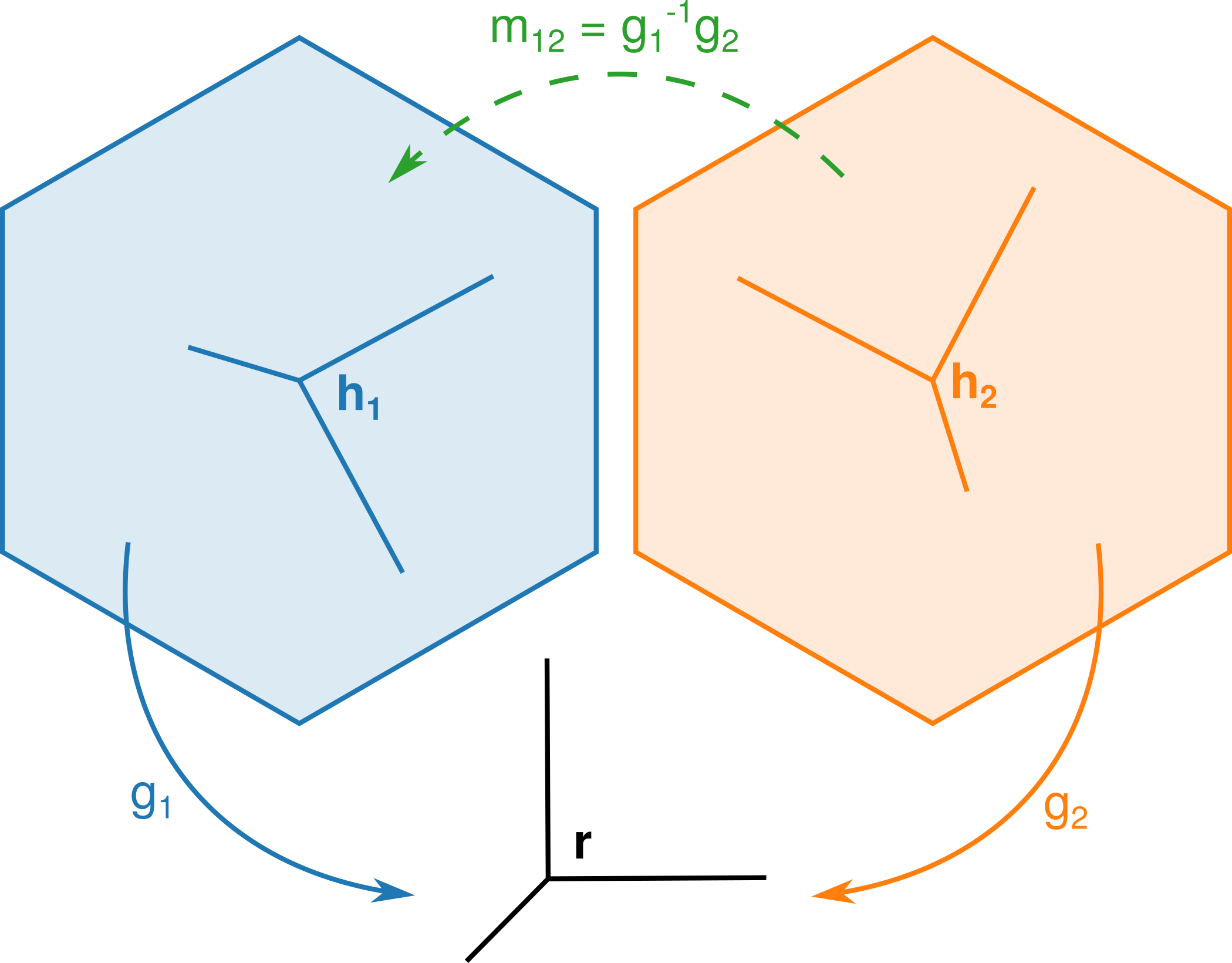}
    \label{fig:misorientation-axisangle}
\end{figure}

Determining the crystal orientation at each two-dimensional pixel or three-dimensional voxel produces a crystal \textit{orientation map}. The \textit{misorientation}, $m$, between crystals at two locations is then the passive rotation between crystal coordinates,

\begin{equation} \label{eq:misorientation}
m_{ij} = g_i^{-1}g_j
\end{equation}

where $g_i$ and $g_j$ are the orientations of each crystal, as illustrated in Figure \ref{fig:misorientation-axisangle}. Since crystal orientations and misorientations are both described as passive rotations in three-dimensions, they can be represented and analysed similarly provided that crystal symmetry is treated appropriately.

Crystal (mis)orientations may be represented as vectors in three-dimensional neo-Eulerian vector spaces based on parametrization of the corresponding axis and angle of rotation \cite{Frank1988,Frank1992}. Visualising (mis)orientation data within the symmetry reduced fundamental zone (or asymmetric domain) of such spaces has recently become more accessible owing to the availability of open-source software packages \cite{Bachmann2010, Groeber2014}. Clusters of (mis)orientations are typically observed within the fundamental zone because (mis)orientation measurements within an individual grain or along a grain boundary are similar. Further, measurements from multiple crystals add to the same cluster if there are preferred crystal orientations or special orientation relationships. Identifying clusters in (mis)orientation data therefore provides a route to identify grains, grain boundaries and orientation relationships. This approach has recently been used to identify grains and crystallographic orientation relationships based on the manual identification of (mis)orientation clusters \cite{Callahan2017, Krakow2017, Krakow2017b, Sunde2019}. However, clusters that cross fundamental zone boundaries appear split due to the crystal symmetry relating the boundaries, which makes the visualisation less clear \cite{Krakow2017}. This motivates a computational approach to (mis)orientation cluster analysis, both to remove manual steps and to improve visualisation.

Clustering of crystal orientations must account for crystal symmetry, which implies that a (mis)orientation is only known up to the action of elements of the proper point group \cite{Krakow2017}. Recently a number of authors have considered the statistics of such ambiguous rotations \cite{Arnold2018, Chen2015a, Niezgoda2016} and hierarchical clustering of (mis)orientations in the presence of crystal symmetry has been demonstrated \cite{Krakow2017c}. Further, a model based clustering algorithm accommodating symmetry, based on a mixture of Von-Mises Fisher or Watson distributions and with parameters estimated using expectation maximization, has also been reported for orientations \cite{Chen2015, Chen2015a}. In this work, we report on density based clustering of (mis)orientations in the presence of crystal symmetry and establish an open-source python library, named \textit{orix}, for handling crystal (mis)orientation data.

%%%%%%%%%%%%%%%%%%%%%%%%%%%%%%%%%%%%%%%%%%%%%%%%%%%%%%%%%%%%%%%%%%%%%%%%%%%
\section{The \textit{orix} python library}\label{sec:method/implementation}
%%%%%%%%%%%%%%%%%%%%%%%%%%%%%%%%%%%%%%%%%%%%%%%%%%%%%%%%%%%%%%%%%%%%%%%%%%%

Orix is a python library designed for the analysis of crystal (mis)orientation data. Here, we describe orix-0.2.0 (released January 2020), which defines various classes and methods that enable: calculations to be performed with three-dimensional rotations; the application of crystal symmetry to rotations for all proper point groups; and the visualization of (mis)orientations in three-dimensional neo-Eulerian vector spaces \cite{Krakow2017}. All rotation calculations are performed in the quaternion representation and conversions between common representations, including Euler angles and axis-angle pairs, are supported \cite{Rowenhorst2015}.

Orix is released open-source \cite{orix} under the GPL-3 license and depends only on core packages in the scientific python stack, namely: \textit{numpy} \cite{numpy}, \textit{scipy} \cite{scipy}, and \textit{matplotlib} \cite{matplotlib}. The code is packaged both on the python package index (PyPI) and the conda-forge repository for use across linux, windows, and mac platforms. A comprehensive set of tests is packaged with the code providing a strong platform for code maintainance and for further development of the package. Usage examples, including the methods described in this paper, are provided online \cite{orix-demos} as a collection of jupyter notebooks \cite{jupyter}.

The development of orix was heavily inspired by the much more extensive Matlab toolbox MTEX \cite{Bachmann2010}. We decided to establish a python library in order to interface more easily with the wider scientific python stack, for example enabling us to directly use clustering algorithms implemented in \textit{scikit-learn} \cite{Pedregosa2011} in this work.

%%%%%%%%%%%%%%%%%%%%%%%%%%%%%%%%%%%%%%%%%%%%%%%%%%%%%%%%%%%%%%
\section{(Mis)orientation clustering method}\label{sec:method}
%%%%%%%%%%%%%%%%%%%%%%%%%%%%%%%%%%%%%%%%%%%%%%%%%%%%%%%%%%%%%%

A cluster analysis is an attempt to partition a set of `objects' $\{o \mid o \in O\}$, such as (mis)orientations, into a meaningful set, $K$, of subsets $\{C \mid C \in K\}$ ($o \in C$), in which the `distance' between objects within each subset $C$ is less than the distance between objects in different subsets \cite{everitt}. To apply this broad definition, a metric for the distance, $d(o_i, o_j)$, between two objects in the set must be defined such that the partition has `meaning' and the conditions $d(o_i, o_i) = 0$ and $d(o_i, o_j) = d(o_j, o_i)$ are satisfied \cite{everitt}. Further, an appropriate clustering algorithm must be selected. Here, distance metrics for (mis)orientations including crystal symmetry and the suitability of density based clustering algorithms for orientation mapping applications are explained.

%%%%%%%%%%%%%%%%%%%%%%%%%%%%%%%%%%%%%%%%%%%%%%%%%%%%%%%%%%%%%%%%%%%%%%%%%%%%%%%%%%%%%%%%%%%%
\subsection{Distance metrics for crystal (mis)orientations}\label{sec:method/misorientations}
%%%%%%%%%%%%%%%%%%%%%%%%%%%%%%%%%%%%%%%%%%%%%%%%%%%%%%%%%%%%%%%%%%%%%%%%%%%%%%%%%%%%%%%%%%%%

The distance, $d(o_i, o_j)$, between two (mis)orientations may be defined as the minimum rotation angle relating them. This rotation angle is symmetric, i.e. it is the same regardless of which orientation is the starting point, and zero for identical (mis)orientations, making it a suitable distance metric for clustering. For crystal (mis)orientations, it is also physical to consider symmetry equivalence. Crystal symmetry implies that the orientation of a crystal with proper point group symmetry, $S$, is equivalent following a transformation $\{s \mid s \in S\}$. This crystal symmetry should be considered in order to determine the minimum rotational angle amongst symmetry equivalent rotations and requires different treatment for orientations and misorientations. An orientation $g$ is equivalent to the set of orientations defined by the equivalence group,

\begin{equation}\label{eq:ori_symmetric}
g = gs, s \in S.
\end{equation}

The rotation between orientations is a misorientation as defined by Equation \ref{eq:misorientation} and combining this definition with Equation \ref{eq:ori_symmetric} yields and expression for symmetrically equivalent misorientations,

\begin{equation}\label{eq:mori_symmetric}
m = s_1ms_2, s_1 \in S_1, s_2 \in S_2
\end{equation}

where $S_1$ and $S_2$ are the symmetry groups of the crystal in each orientation.

The distance between two orientations, $g_i$ and $g_j$, associated with crystals with the symmetry groups $S_k$ and $S_l$ respectively is thus given by,

\begin{equation}\label{eq:ori_dist}
d(g_i, g_j) = \min_{s_k \in S_k} s_k m s_l
\end{equation}

The distance between two misorienations is defined similarly as the rotation between two misorientations, $m_i$ to $m_j$, which, accounting for the crystal symmetry of the two pairs of crystals associated with each misorientation using Equation \ref{eq:mori_symmetric}, gives,

\begin{equation}\label{eq:mori_dist}
d(m_{i}, m_{j}) = \min_{s_k \in S_k} s_k m_{i}^{-1} s_l s_q m_{j} s_r
\end{equation}

where $i, j$ are indices indicating (mis)orientations associated with an orientation map and $k, l, q, r$ are indices indicating the symmetry group corresponding to the the crystal phase associated with each (mis)orientation.

%%%%%%%%%%%%%%%%%%%%%%%%%%%%%%%%%%%%%%%%%%%%%%%%%%%%%%%%%%%%%%%%%%%%%%%%%
\subsection{Density-based clustering of (mis)orientations}\label{sec:method/computation}
%%%%%%%%%%%%%%%%%%%%%%%%%%%%%%%%%%%%%%%%%%%%%%%%%%%%%%%%%%%%%%%%%%%%%%%%%

A distance matrix, $D_{ij}$, containing the distance between all (mis)orientations, may be defined using Equations \ref{eq:ori_dist} \& \ref{eq:mori_dist} and used to initialise a clustering algorithm. In clustering (mis)orientation data, we aim to identify an unknown number of small dense clusters associated with grains, grain boundaries and special orientation relationships while excluding spurious data points resulting from incorrect automated indexation. Density-based clustering methods are well suited to this application because they are based on identifying clusters as regions of higher density than the remainder of the dataset while identifying points in sparse regions as noise or boundary points. This contrasts with centroid- and model- based methods that typically require a good estimate of the number of clusters and hierarchical clustering, which does not provide a unique partition and is not very robust to outliers \cite{everitt}. We note that model-based and hierarchical clustering methods have nevertheless been demonstrated to provide useful (mis)orientation clustering \cite{Chen2015a, Krakow2017c}.

We perform density-based clustering using the DBSCAN algorithm \cite{Ester1996} implemented in \textit{scikit-learn} \cite{Pedregosa2011}. Two parameters are required: \(\epsilon\) - an angle (in radians) acting as the upper limit on the distance between two points to be considered to be in the neighborhood of one another; and $n$ - the minimum number of data points in the core of a cluster. Because the DBSCAN algorithm is reasonably quick it is possible to cluster with a range of parameters. We obtained reasonable results using \(\epsilon = 0.05\) and \(n = 40\) for orientations and \(\epsilon = 0.05\), \(n=10\) for misorientations.

%%%%%%%%%%%%%%%%%%%%%%%%%%%%%%%%%%%%%%%%%%%%%
\section{(Mis)orientation clustering results}
%%%%%%%%%%%%%%%%%%%%%%%%%%%%%%%%%%%%%%%%%%%%%

An orientation map obtained via EBSD mapping of a commercially pure hexagonal close packed (hcp) titanium (point group 6/mmm) sample, following high strain rate deformation, was used to illustrate the density-based (mis)orientation clustering method. This dataset was downloaded from an online repository \cite{Krakow2017a} for this demonstration and was previously described in detail by \citeasnoun{Krakow2017}. The orientation map contains data from two parent grains, each containing deformation twins.

%%%%%%%%%%%%%%%%%%%%%%%%%%%%%%%%%%%%%%%%%%%%%%%%%%%
\subsection{Clustering orientations to find grains}
%%%%%%%%%%%%%%%%%%%%%%%%%%%%%%%%%%%%%%%%%%%%%%%%%%%

\begin{figure}
    \caption{(a) Crystal orientations plotted within the fundamental zone for symmetry group \textit{622} in axis-angle space and coloured to indicate cluster membership as determined using the DBSCAN algorithm. Axes are labelled in the crystallographic basis at no rotation. (b) Map of a twinned Ti microstructure coloured by cluster membership of the orientation associated with each pixel.}
    \includegraphics[]{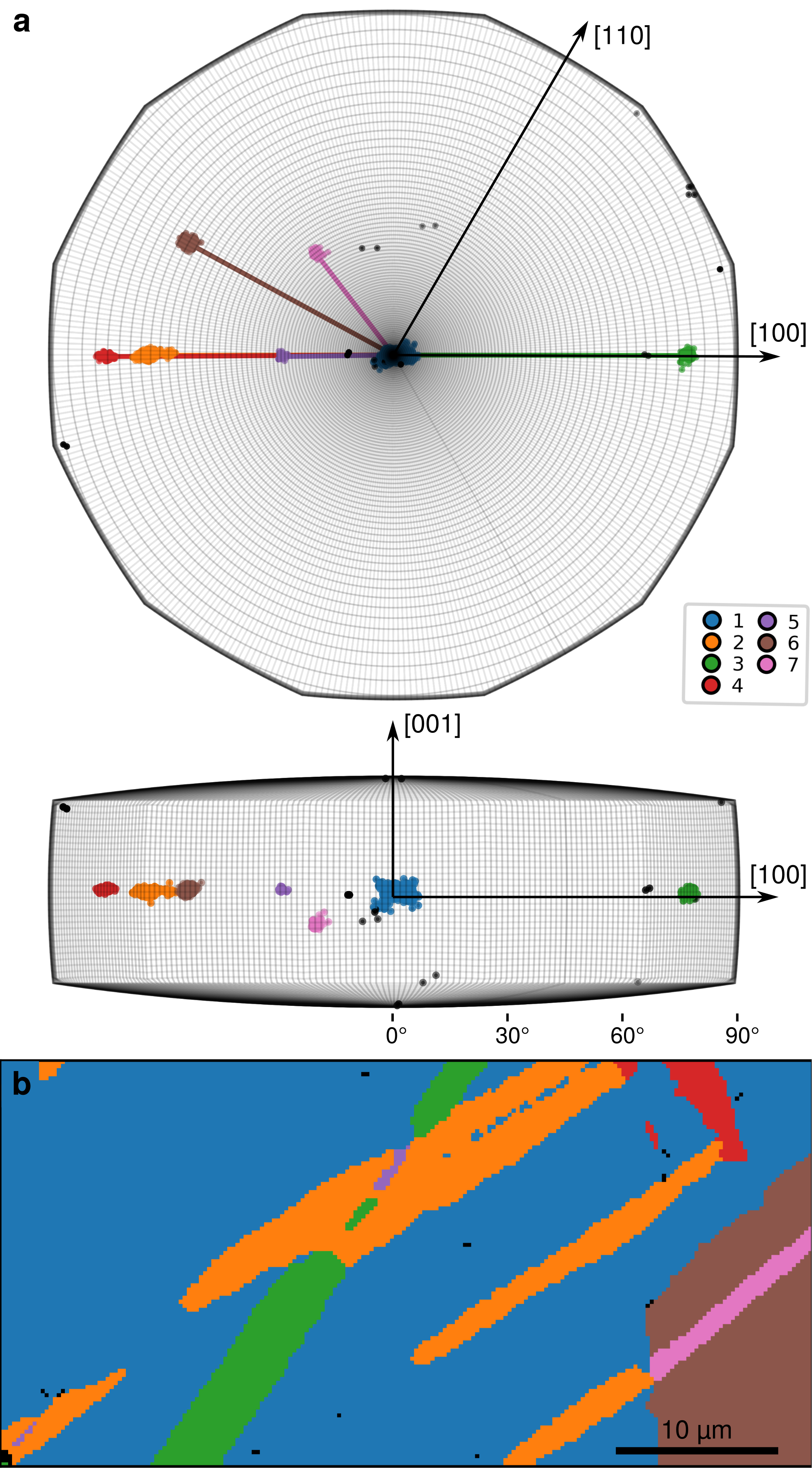}
    \label{fig:clustered-orientations}
\end{figure}

Orientation clusters, determined by density-based clustering of the data, are shown in Figure \ref{fig:clustered-orientations}(a). The clusters are plotted within the asymmetric domain of axis-angle space \cite{Krakow2017} for the proper point group, \textit{622}, of hcp titanium and the mean orientation of the largest parent grain (cluster 1) is taken as the reference orientation. Clusters 2-5 are all rotated about $[100]$ with respect to the reference parent grain (cluster 1) suggesting that they may correspond to twins, whereas clusters 6 and 7 are rotated about other axes.

Plotting the spatial location associated with data points in each orientation cluster, as shown in Figure \ref{fig:clustered-orientations}(b), provides a clear visualisation of the grain structure and illustrates that the clustering result is physically meaningful. Clusters 2-5 correspond to lenticular grains, typical of deformation twins, within the larger parent grain (cluster 1). Cluster 6 corresponds to the second parent grain and cluster 7 to a lenticular deformation twin within that grain. Some data points are not assigned to any cluster and correspond to misindexed pixels. We note that despite the asymmetrical shape of some clusters (eg. clusters 1 and 2) resulting from deformation within the grain this has not caused issues with this clustering.

%%%%%%%%%%%%%%%%%%%%%%%%%%%%%%%%%%%%%%%%%%%%%%%%%%%%%%%%%%%
\subsection{Clustering misorientations at grain boundaries}
%%%%%%%%%%%%%%%%%%%%%%%%%%%%%%%%%%%%%%%%%%%%%%%%%%%%%%%%%%%

\begin{figure}
    \caption{(a) Crystal misorientations plotted in the fundamental zone for the symmetry group pair (\textit{622}, \textit{622)} in axis-angle space and coloured to indicate cluster membership as determined using the DBSCAN algorithm. Axes are labelled in the crystallographic basis at no rotation. (b) Map of grain boundaries coloured by cluster membership of the misorientation at with each boundary element.}
    \includegraphics[]{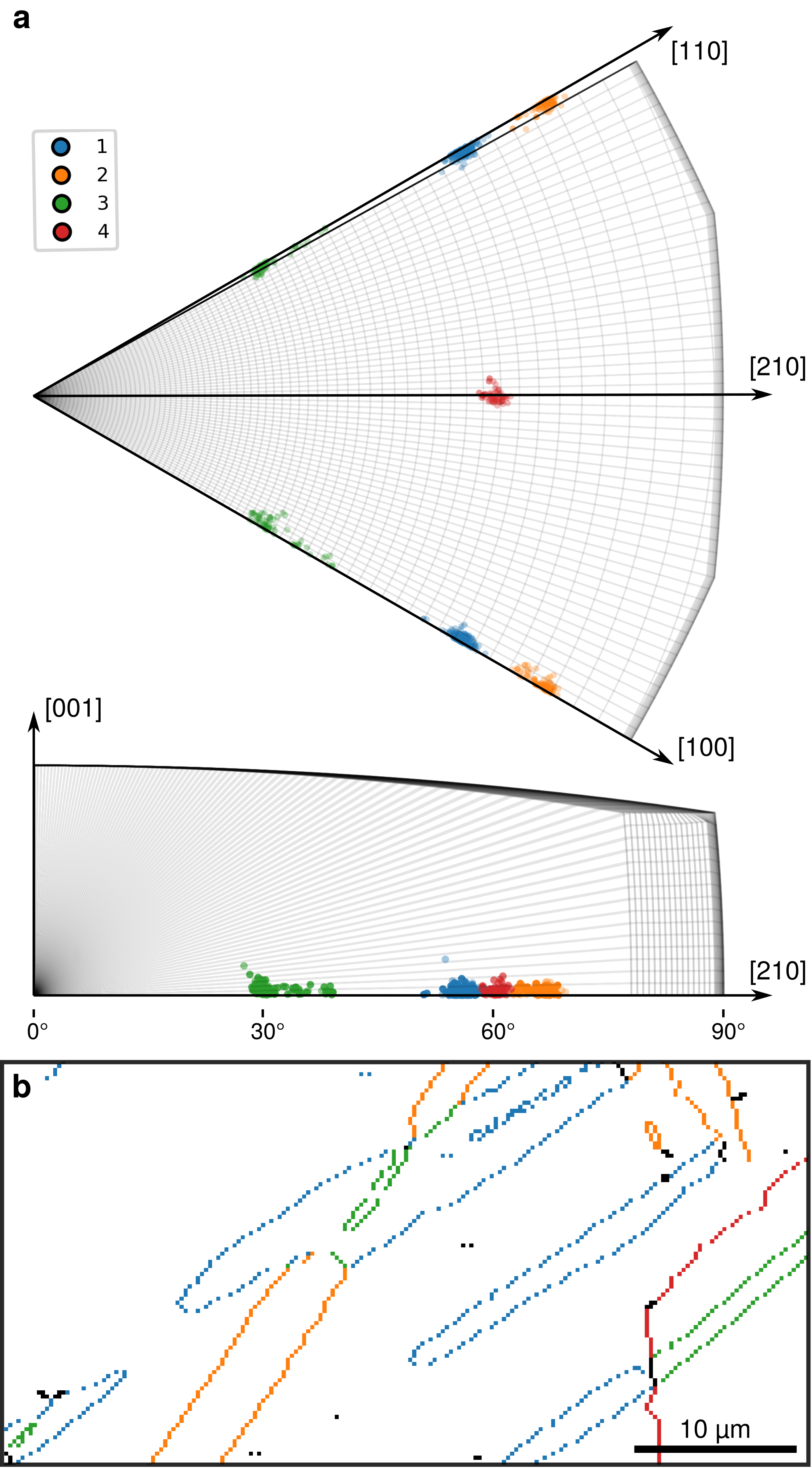}
    \label{fig:clustered-misorientations}
\end{figure}

The misorientation between horizontally adjacent pixels was computed from the orientation mapping data. Misorientations with rotation angles less than \SI{7}{\degree}, corresponding to the grain orientation spread within the largest grain in this highly deformed material, were discarded in order to identify grain boundaries. Misorientation clusters, determined by density-based clustering of this data are shown in Figure \ref{fig:clustered-misorientations}(a). These misorientations are plotted within the asymmetric domain of axis-angle space for misorientations between two hcp titanium crystals each with proper point group symmetry, \textit{622}, without application of grain exchange symmetry \cite{Krakow2017}. Four clusters are identified, three of which (clusters 1-3) are situated across the boundary of the asymmetric domain and are identified as belonging to the same cluster owing to the inclusion of crystal symmetry in the distance metric.

The mean misorientation associated with each cluster highlighted in Figure \ref{fig:clustered-misorientations}(a) was calculated as the quaternion mean \cite{Morawiec1998} of misorientations in the cluster. The minimum rotational angle between these cluster centers and theoretical misorientations associated with near coincident site lattice (n-CSL) orientation relationships \cite{Bonnet1981}, which result from deformation twinning \cite{Laine2015}, were computed to determine the closest n-CSL to each cluster centre. Clusters 1-3 were found to be within ca. $1.2\,^{\circ}$ of n-CSL relationships associated with deformation twinning, whereas cluster 4 was $\>4\,^{\circ}$ from the nearest n-CSL relationship, as reported in Table \ref{tab:mori_comparison}. This suggests that clusters 1-3 correspond to deformation twin boundaries, whereas cluster 4 does not. Inspecting the spatial distribution of misorientation clusters, as in Figure \ref{fig:clustered-misorientations}, confirms that clusters 1-3 correspond to deformation twin boundaries, whereas cluster 4 corresponds to the boundary between parent grains. All remaining points correspond to misindexed pixels.

\begin{table}
    \centering
    \label{tab:mori_comparison}
    \caption{Comparison of misorientation cluster mean values with near coincident site lattice (n-CSL) misorientations \cite{Bonnet1981} calculated for titanium with an assumed c/a=1.588. }
    \begin{tabular}{cccc}      % Alignment for each cell: l=left, c=center, r=right
    cluster & nearest n-CSL & theoretical misorientation & distance \\
    1 & n-CSL7a  & $[100]$ 64.40$^{\circ}$ & $0.44\,^{\circ}$ \\
    2 & n-CSL13a & $[100]$ 76.89$^{\circ}$ & $0.70\,^{\circ}$ \\
    3 & n-CSL11a & $[100]$ 34.96$^{\circ}$ & $1.19\,^{\circ}$ \\
    4 & n-CSL13b & $[210]$ 57.22$^{\circ}$ & $4.44\,^{\circ}$ \\
    \end{tabular}
\end{table}

%%%%%%%%%%%%%%%%%%%%
\section{Discussion}
%%%%%%%%%%%%%%%%%%%%

Density-based clustering using a distance metric that accounts for crystal symmetry has been demonstrated above to successfully identify grain structure and grain boundary structure in experimental orientation mapping data. This includes treatment of spurious misindexed pixels and elongated asymmetrical clusters due to distortions within grains. The DBSCAN algorithm used requires only two parameters to be set and therefore minimal prior knowledge. The clustering results enhance the practical utility of three-dimensional misorientation spaces as a tool for investigating orientation mapping data by automatically identifying clusters. In particular, clusters that cross the boundaries of the asymmetric domain are identified and can be indicated when plotting the data, making visualizations easier to interpret. Plotting the spatial distribution of (mis)orientation clusters further provides an easy way to relate observations in real space and (mis)orientation space.

The analysis is not without limitations. Density-based clustering algorithms are known to struggle with datasets in which the overall density is high as a density drop is needed to identify cluster boundaries. This could occur when an orientation map contains data from a large number of grains and in such cases an alternative solution may be more suitable, for example recently reported model based clustering \cite{Chen2015, Chen2015a} or hierarchical clustering \cite{Krakow2017c}. A further limitation is that the clustering labels clusters but does not estimate any parameters associated with the (mis)orientation distribution. Using the cluster centers as a starting point for fitting local (mis)orientation distribution functions may therefore be an important extension. Another challenge lies in identifying whether or not a cluster constitutes a statistically significant observation with respect to finite sampling of a random distribution, which merits further investigation.

%%%%%%%%%%%%%%%%%%%%%
\section{Conclusions}
%%%%%%%%%%%%%%%%%%%%%

This work demonstrates that density based clustering of crystal orientations and misorientations, using a distance metric accounting for crystal symmetry and the DBSCAN algorithm, can provide important physical insights using very little prior knowledge. In particular, we used this approach to identify characteristic misorientations associated with deformation twinning as an illustrative example of how the approach may be used to identify special orientation relationships. A python library, named \textit{orix} was established to provide various classes and methods required for the manipulation of (mis)orientation data and it is hoped that this library will serve as a platform for further developments.

     % Appendices appear after the main body of the text. They are prefixed by
     % a single \appendix declaration, and are then structured just like the
     % body text.

% \appendix
% \section{Appendix title}

% Text text text text text text text text text text text text text text
% text text text text text text text.

% \subsection{Title}

% Text text text text text text text text text text text text text text
% text text text text text text text.

% \subsubsection{Title}

% Text text text text text text text text text text text text text text
% text text text text text text text.

     %-------------------------------------------------------------------------
     % The back matter of the paper - acknowledgements and references
     %-------------------------------------------------------------------------

     % Acknowledgements come after the appendices

\ack{Acknowledgements}

The authors acknowledge financial support from The Royal Society and the EPSRC under grant number EP/R008779/1 and the studentship 1937212 in partnership with the National Physical Laboratory. The authors would also like to thank Dr Robert Krakow for discussions that initiated this work.

     % References are at the end of the document, between \begin{references}
     % and \end{references} tags. Each reference is in a \reference entry.

% \begin{references}
% \reference{Author, A. \& Author, B. (1984). \emph{Journal} \textbf{Vol}, 
% first page--last page.}
% \end{references}
% \cite{knuth84}

\bibliographystyle{iucr}

@software{orix,
  author       = {Ben Martineau and
                  Phillip Crout and
                  Duncan N. Johnstone},
  title        = {pyxem/orix: orix 0.2.0},
  month        = dec,
  year         = 2019,
  publisher    = {Zenodo},
  version      = {v0.2.0},
  doi          = {10.5281/zenodo.3598729},
  url          = {https://doi.org/10.5281/zenodo.3598729}
}

@software{orix-demos,
  author       = {Duncan N. Johnstone and
                  Phillip Crout},
  title        = {pyxem/orix-demos: orix-demos 0.2.0},
  month        = dec,
  year         = 2019,
  publisher    = {Zenodo},
  version      = {v0.2.0},
  doi          = {10.5281/zenodo.3600315},
  url          = {https://doi.org/10.5281/zenodo.3600315}
}

@article{numpy,
  author    = {St{\'{e}}fan van der Walt and
               S. Chris Colbert and
               Ga{\"{e}}l Varoquaux},
  title     = {The NumPy array: a structure for efficient numerical computation},
  journal = {arXiv e-prints},
  year      = {2011},
  pages     = {arXiv:1102.1523}
}

@conference{jupyter,
	Author = {Thomas Kluyver and Benjamin Ragan-Kelley and Fernando P{\'e}rez and Brian Granger and Matthias Bussonnier and Jonathan Frederic and Kyle Kelley and Jessica Hamrick and Jason Grout and Sylvain Corlay and Paul Ivanov and Dami{\'a}n Avila and Safia Abdalla and Carol Willing},
	Booktitle = {Positioning and Power in Academic Publishing: Players, Agents and Agendas},
	Editor = {F. Loizides and B. Schmidt},
	Organization = {IOS Press},
	Pages = {87 - 90},
	Title = {Jupyter Notebooks -- a publishing format for reproducible computational workflows},
	Year = {2016}}

@ARTICLE{scipy,
       author = {{Virtanen}, Pauli and {Gommers}, Ralf and {Oliphant},
         Travis E. and {Haberland}, Matt and {Reddy}, Tyler and
         {Cournapeau}, David and {Burovski}, Evgeni and {Peterson}, Pearu
         and {Weckesser}, Warren and {Bright}, Jonathan and {van der Walt},
         St{\'e}fan J.  and {Brett}, Matthew and {Wilson}, Joshua and
         {Jarrod Millman}, K.  and {Mayorov}, Nikolay and {Nelson}, Andrew
         R.~J. and {Jones}, Eric and {Kern}, Robert and {Larson}, Eric and
         {Carey}, CJ and {Polat}, {\.I}lhan and {Feng}, Yu and {Moore},
         Eric W. and {Vand erPlas}, Jake and {Laxalde}, Denis and
         {Perktold}, Josef and {Cimrman}, Robert and {Henriksen}, Ian and
         {Quintero}, E.~A. and {Harris}, Charles R and {Archibald}, Anne M.
         and {Ribeiro}, Ant{\^o}nio H. and {Pedregosa}, Fabian and
         {van Mulbregt}, Paul and {Contributors}, SciPy 1. 0},
        title = "{SciPy 1.0--Fundamental Algorithms for Scientific
                  Computing in Python}",
      journal = {arXiv e-prints},
         year = "2019",
        month = "Jul",
          eid = {arXiv:1907.10121},
        pages = {arXiv:1907.10121},
archivePrefix = {arXiv},
       eprint = {1907.10121},
 primaryClass = {cs.MS},
       adsurl = {https://ui.adsabs.harvard.edu/abs/2019arXiv190710121V}
}

@Article{matplotlib,
  Author    = {Hunter, J. D.},
  Title     = {Matplotlib: A 2D graphics environment},
  Journal   = {Computing in Science \& Engineering},
  Volume    = {9},
  Number    = {3},
  Pages     = {90--95},
  publisher = {IEEE COMPUTER SOC},
  doi       = {10.1109/MCSE.2007.55},
  year      = 2007
}

@article{Arnold2018,
title = "Statistics of ambiguous rotations",
journal = "Journal of Multivariate Analysis",
volume = "165",
pages = "73 - 85",
year = "2018",
issn = "0047-259X",
doi = "https://doi.org/10.1016/j.jmva.2017.10.007",
author = "R. Arnold and P.E. Jupp and H. Schaeben"
}

@inproceedings{Bachmann2010,
author = {Bachmann, F. and Hielscher, Ralf and Schaeben, Helmut},
title = {Texture Analysis with MTEX – Free and Open Source Software Toolbox},
year = {2010},
month = {3},
volume = {160},
pages = {63--68},
booktitle = {Texture and Anisotropy of Polycrystals III},
series = {Solid State Phenomena},
publisher = {Trans Tech Publications},
doi = {10.4028/www.scientific.net/SSP.160.63}
}

@article{Callahan2017,
author = "Callahan, Patrick G. and Echlin, McLean and Pollock, Tresa M. and Singh, Saransh and De Graef, Marc",
title = "{Three-dimensional texture visualization approaches: theoretical analysis and examples}",
journal = "Journal of Applied Crystallography",
year = "2017",
volume = "50",
number = "2",
pages = "430--440",
month = "Apr",
doi = {10.1107/S1600576717001157},
}

@INPROCEEDINGS{Chen2015,
author={Y. Chen and D. Wei and G. Newstadt and M. DeGraef and J. Simmons and A. Hero},
booktitle={18th International Conference on Information Fusion (Fusion)},
title={Statistical estimation and clustering of group-invariant orientation parameters},
year={2015},
pages={719-726},
month={July}
}

@ARTICLE{Chen2015a,
author={Y. Chen and D. Wei and G. Newstadt and M. DeGraef and J. Simmons and A. Hero},
journal={IEEE Signal Processing Letters},
title={Parameter Estimation in Spherical Symmetry Groups},
year={2015},
volume={22},
number={8},
pages={1152-1155},
doi={10.1109/LSP.2014.2387206},
month={Aug}
}

@INPROCEEDINGS{Ester1996,
author = {Martin Ester and Hans-Peter Kriegel and Jörg Sander and Xiaowei Xu},
title = {A density-based algorithm for discovering clusters in large spatial databases with noise},
booktitle = {},
year = {1996},
pages = {226--231},
publisher = {AAAI Press}
}

@book{Everitt,
title={Cluster Analysis},
author={Everitt, B.S. and Landau, S. and Leese, M. and Stahl, D.},
isbn={9780470978443},
lccn={2010037932},
series={Wiley Series in Probability and Statistics},
year={2011},
publisher={Wiley}
}

@article{Bonnet1981,
author = "Bonnet, R. and Cousineau, E. and Warrington, D. H.",
title = "{Determination of near-coincident cells for hexagonal crystals. Related DSC lattices}",
journal = "Acta Crystallographica Section A",
year = "1981",
volume = "37",
number = "2",
pages = "184--189",
month = "Mar"
}

@article{Laine2015,
author = {Steven J. Lainé and Kevin M. Knowles},
title = {{11¯24} deformation twinning in commercial purity titanium at room temperature},
journal = {Philosophical Magazine},
volume = {95},
number = {20},
pages = {2153-2166},
year  = {2015}
}

@article{Frank1988,
title = {{Orientation mapping}},
year = {1988},
journal = {Metallurgical Transactions A},
author = {Frank, F. C.},
number = {3},
pages = {403--408},
volume = {19},
isbn = {9780444537706},
doi = {10.1007/BF02649253},
issn = {0360-2133}
}

@article{Frank1992,
title = {{The conformal neo-eulerian orientation map}},
year = {1992},
journal = {Philosophical Magazine A},
author = {Frank, F C},
number = {5},
pages = {1141--1149},
volume = {65},
isbn = {0141861920820},
doi = {10.1080/01418619208201501},
issn = {01418610}
}

@Article{Groeber2014,
author="Groeber, Michael A.
and Jackson, Michael A.",
title="DREAM.3D: A Digital Representation Environment for the Analysis of Microstructure in 3D",
journal="Integrating Materials and Manufacturing Innovation",
year="2014",
month="Apr",
day="01",
volume="3",
number="1",
pages="5",
doi="10.1186/2193-9772-3-5"
}

@article{Ice2009,
title = "Tutorial on x-ray microLaue diffraction",
journal = "Materials Characterization",
volume = "60",
number = "11",
pages = "1191 - 1201",
year = "2009",
issn = "1044-5803",
doi = "https://doi.org/10.1016/j.matchar.2009.07.006",
author = "Gene E. Ice and Judy W.L. Pang"
}

@book{Kocks1998,
title = {{Texture and Anisotropy - Preferred Orientations in Polycrystals and their Effect on Materials Properties}},
year = {1998},
author = {Kocks, U.F. and Tome, C.N. and Wenk, H.-R.},
pages = {676},
publisher = {Cambridge University Press}
}

@article {Krakow2017,
author = {Krakow, Robert and Bennett, Robbie J. and Johnstone, Duncan N. and Vukmanovic, Zoja and Solano-Alvarez, Wilberth and Lain{\'e}, Steven J. and Einsle, Joshua F. and Midgley, Paul A. and Rae, Catherine M. F. and Hielscher, Ralf},
title = {On three-dimensional misorientation spaces},
volume = {473},
number = {2206},
year = {2017},
doi = {10.1098/rspa.2017.0274},
publisher = {The Royal Society},
issn = {1364-5021},
eprint = {http://rspa.royalsocietypublishing.org/content/473/2206/20170274.full.pdf},
journal = {Proceedings of the Royal Society of London A: Mathematical, Physical and Engineering Sciences}
}

@misc{Krakow2017a,
title = {{MTEX scripts and EBSD supporting 'On Three-dimensional Misorientation Spaces' https://doi.org/10.17863/CAM.8815 .}},
year = {2017},
author = {Krakow, R. and Hielscher, R.},
pages = {1--10},
url = {https://github.com/mtex-toolbox/mtex-paper/tree/master/3dMisorientationSpace},
doi = {https://doi.org/10.17863/CAM.8815}
}

@article{Krakow2017b,
title = "On the crystallography and composition of topologically close-packed phases in ATI 718Plus®",
journal = "Acta Materialia",
volume = "130",
pages = "271 - 280",
year = "2017",
issn = "1359-6454",
author = "Robert Krakow and Duncan N. Johnstone and Alexander S. Eggeman and Daniela Hünert and Mark C. Hardy and Catherine M.F. Rae and Paul A. Midgley"
}

@article{Krakow2017c,
title={Inter-phase Relationships Revealed in 3-Dimensional Orientation Spaces},
volume={23},
DOI={10.1017/S1431927617001696},
number={S1},
journal={Microscopy and Microanalysis},
publisher={Cambridge University Press},
author={Krakow, Robert and Bennett, Robbie J. and Johnstone, Duncan N. and Midgley, Paul A. and Hielsher, Ralf and Rae, Catherine M. F.},
year={2017},
pages={202–203}
}

@article{Morawiec1998,
author = "Morawiec, A.",
title = "{A note on mean orientation}",
journal = "Journal of Applied Crystallography",
year = "1998",
volume = "31",
number = "5",
pages = "818--819",
month = "Oct",
doi = {10.1107/S0021889898003914}
}

@book{Morawiec2004,
author    = {Adam Morawiec}, 
title     = {Orientations and Rotations},
publisher = {Springer-Verlag},
year      = 2004,
address   = {Berlin},
edition   = 1,
isbn      = {978-3-662-09156-2}
}

@article{Niezgoda2016,
author = "Niezgoda, Stephen R. and Magnuson, Eric A. and Glover, Jared",
title = "{Symmetrized Bingham distribution for representing texture: parameter estimation with respect to crystal and sample symmetries}",
journal = "Journal of Applied Crystallography",
year = "2016",
volume = "49",
number = "4",
pages = "1315--1319",
month = "Aug",
doi = {10.1107/S160057671600649X}
}

@article{Pedregosa2011,
 title={Scikit-learn: Machine Learning in {P}ython},
 author={Pedregosa, F. and Varoquaux, G. and Gramfort, A. and Michel, V.
         and Thirion, B. and Grisel, O. and Blondel, M. and Prettenhofer, P.
         and Weiss, R. and Dubourg, V. and Vanderplas, J. and Passos, A. and
         Cournapeau, D. and Brucher, M. and Perrot, M. and Duchesnay, E.},
 journal={Journal of Machine Learning Research},
 volume={12},
 pages={2825--2830},
 year={2011}
}

@article{Rauch2008,
title = {{Automatic Crystal Orientation and Phase Mapping in TEM by Precession Diffraction}},
year = {2008},
journal = {Microscopy and Analysis Nanotechnolog Supplement},
author = {Rauch, E F and V{\'{e}}ron, M and Portillo, Joaquim and Bultreys, Daniel and Maniette, Y and Nicolopoulos, Stavros},
number = {6},
pages = {5--8},
volume = {22}
}

@article{Rowenhorst2015,
title = {{Consistent representations of and conversions between 3D rotations}},
year = {2015},
journal = {Modelling and Simulation in Materials Science and Engineering},
author = {Rowenhorst, D and Rollett, A D and Rohrer, G S and Groeber, M and Jackson, M and Konijnenberg, P J and De Graef, M},
number = {8},
pages = {083501},
volume = {23},
doi = {10.1088/0965-0393/23/8/083501}
}

@book{Schwartz2009,
title = {{Electron Backscatter Diffraction in Materials Science}},
year = {2009},
author = {Schwartz, A.J., Kumar, M., Adams, B.L., Field, D.P. (Eds.)},
pages = {403},
publisher = {Springer},
isbn = {978-0-387-88135-5}
}

@article{Sunde2019,
title = "Crystallographic relationships of T-/S-phase aggregates in an Al–Cu–Mg–Ag alloy",
journal = "Acta Materialia",
volume = "166",
pages = "587 - 596",
year = "2019",
issn = "1359-6454",
doi = "https://doi.org/10.1016/j.actamat.2018.12.036",
author = "Jonas K. Sunde and Duncan N. Johnstone and Sigurd Wenner and Antonius T.J. van Helvoort and Paul A. Midgley and Randi Holmestad"
}

@book{Sutton2007,
title = {{Interfaces in Crystalline Materials}},
year = {2007},
author = {Sutton, A. P. and Baluffi, R.W.},
pages = {856},
publisher = {Oxford University Press},
address = {Oxford},
isbn = {019921106X}
}

@article{Zaefferer2000,
author = "Zaefferer, Stefan",
title = "{New developments of computer-aided crystallographic analysis in transmission electron microscopy}",
journal = "Journal of Applied Crystallography",
year = "2000",
volume = "33",
number = "1",
pages = "10--25",
month = "Feb",
doi = {10.1107/S0021889899010894}
}
\referencelist

     %-------------------------------------------------------------------------
     % TABLES AND FIGURES SHOULD BE INSERTED AFTER THE MAIN BODY OF THE TEXT
     %-------------------------------------------------------------------------

\end{document}